\documentclass[twocolumn,letterpaper,aps,prl,showpacs,floatfix,10pt]{revtex4-1}
\usepackage{graphicx}    
\usepackage{hyperref}    
\usepackage{bm}          
\usepackage[sumlimits,intlimits]{amsmath}
\usepackage{amsfonts,amssymb}
\pdfoutput=1
\begin{document}

\title{Apparent power-law behavior of conductance in disordered quasi-one-dimensional systems}
\author{A. S. Rodin}
\author{M. M. Fogler}
\affiliation{University of California San Diego, 9500 Gilman Drive, La
Jolla, California 92093}

\date{\today}

\begin{abstract}
Dependence of hopping conductance on temperature and voltage for an ensemble of modestly long one-dimensional wires is studied numerically using the shortest-path algorithm. In a wide range of parameters this dependence can be approximated by a power law rather than the usual stretched-exponential form. Relation to recent experiments and prior analytical theory is discussed.
\end{abstract}


\maketitle

Localization by disorder is a common cause of insulating behavior of low-dimensional electron systems. If the system size greatly exceeds the localization length, the transport at low voltages $V$ and temperatures $T$ is governed by the variable-range hopping (VRH)~\cite{Shklovskii1984epo}. The electric current $I(V, T)$ and the conductance $G(V, T) \equiv I(V, T) / V$ have a (stretched) exponential behavior. For example, in the Ohmic regime 
\begin{equation}\label{eqn:VRH}
 G_{\Omega} \equiv G(0, T) \propto \exp \left[-(T_* / T)^\mu \right]\,,
 \quad
 0 < \mu \leq 1\,.
\end{equation}
Over the last decade, observations of different laws,
\begin{align}
I &\propto V T^{\alpha},
\quad V \ll (2 \pi / \gamma) T,
\label{eqn:alpha}\\
&\propto V^{\beta + 1},
\quad V \gg (2 \pi / \gamma) T,
\label{eqn:beta}
\end{align}
have been reported in systems as diverse as carbon nanotubes~\cite{Bockrath1999llb, Yao1999cni, Bachtold2001sot, Kanda2004gvd, Gao2004efl, Monteverde2006tll, Coiffic2007cri, Dayen2009cod}, InSb~\cite{Zaitsev2000lll} and GaAs~\cite{Chang2003cll, Tserkovnyak2003iaz} quantum wires, NiSe$_3$ whiskers~\cite{Slot2004odc},  polymer nanofibers~\cite{Aleshin2004odt, Rahman2010bdc}, inorganic~\cite{Venkataraman2006eti, Dayen2009cod} and organic nanowires~\cite{Zhou2007ode}, as well as polymer films~\cite{Yuen2009nti, Worne2010tio}. The coefficients $\alpha$, $\beta$, and $\gamma$ vary among different materials and different samples of the same material.

A five-parameter formula frequently used to fit the experimental data is
\begin{equation}
I = c_0 T^{\alpha + 1}
          \sinh \left(\frac{\gamma^\prime V}{2 T} \right)
          \left| \Gamma\left(1 + \frac{\beta}{2} + i\, \frac{\gamma V}{2\pi T}\right)\right|^2.
\label{eqn:LL}
\end{equation}
For $\gamma^\prime = \gamma$ the asymptotic behavior of $I(V, T)$ is given by Eqs.~\eqref{eqn:alpha} and \eqref{eqn:beta}. Agreement with Eq.~\eqref{eqn:LL} was advocated as evidence for tunneling into Luttinger liquid (LL)~\cite{Giamarchi2004qpi} --- a one-dimensional (1D) system with nonperturbative interaction effects. (For strong interactions the LL can also be modeled as a 1D Wigner crystal~\cite{[{For a recent review, see }]Deshpande2010ela}.) In this picture, the system contains a tunneling barrier, e.g., a poor contact, but is otherwise clean and free of localization. The power-laws are due to renormalization of this barrier by many-body effects. However, there is a problem with this interpretation. The actual calculations~\cite{Kane1992tia, Sassetti1994to1, Balents1999oci, Mishchenko2001zba} within the LL model give $\alpha = \beta$ and $\gamma = \gamma^\prime = 1$, which is not always consistent with the parameters of the empirical fits (a notable exception is Ref.~\onlinecite{Chang2003cll}).

Another reason to doubt the relevance of the LL effects in some of these experiments is the fact that the systems studied are neither perfectly clean nor truly 1D. They are, typically, collections of many parallel 1D channels, whose total number ranges from several hundred to many thousands, each containing multiple impurities.

In this Letter we show that in such quasi-1D systems the conventional mechanism of transport, which is the VRH, can also lead to Eqs.~\eqref{eqn:alpha}--\eqref{eqn:LL}. This is because at low enough $T$ the hopping length is not much smaller than the length $L$ of the wires. In this case, the VRH conductance deviates from the usual formula, Eq.~\eqref{eqn:VRH}. The hopping is dominated by hopping paths that consist of a few approximately equidistant hops~\cite{Tartakovskii1987hco, Levin1988thc, Glazman1988ita, Bahlouli1994cci}. Although rare, such configurations can always be found in a sample if the number of channels is large enough. Hence, despite mesoscopic fluctuations that accompany rare events, $G(V, T)$ can be a smooth quasi-power-law function.

%
%
\begin{figure}[b]
  \includegraphics[width=3in]{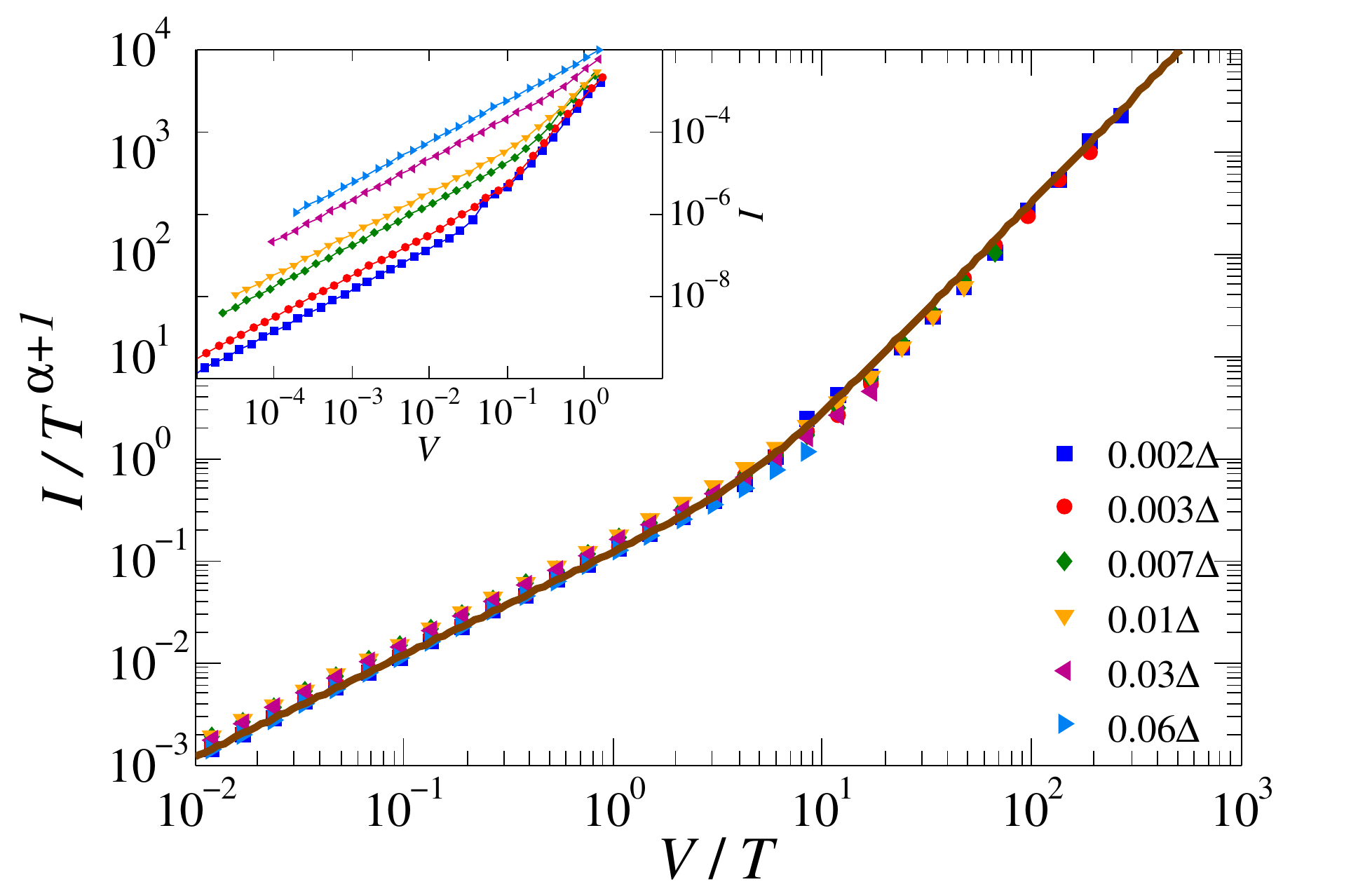}
  \caption{\label{fig:L30_Collapse} (Color online) Main panel: Collapse onto the ``universal curve'' of Eq. (4) (solid line) obtained by plotting the numerical results for $L = 30$ and $a = 4$ (symbols) as $I / T^{\alpha +1}$ \textit{vs}. $V/T$, using $\alpha = 1.75$, $\beta = 1.1$, and $\gamma = 1$. The temperatures are listed in the legend in the units of $\Delta = 4 T_0$. Inset: Same data plotted as $I$ \textit{vs}. $V$.}
\end{figure}

Our results are illustrated in Fig.~\ref{fig:L30_Collapse}. They are obtained numerically following the approach used in our previous work~\cite{Rodin2009nso}, with some improvements. In the inset of Fig.~\ref{fig:L30_Collapse} we show a set of $I$-$V$ curves computed for a set of fixed $T$. In the main panel, we collapse them onto a single ``universal'' curve described by Eq.~\eqref{eqn:LL}.

Let us compare the quality of the data collapse to those in the aforementioned experiments~\cite{Venkataraman2006eti, Slot2004odc, Aleshin2004odt, Zhou2007ode, Rahman2010bdc, Coiffic2007cri, Yuen2009nti, Worne2010tio}. The range of $T$ shown in Fig.~\ref{fig:L30_Collapse} is a factor of $30$. In the experiments, Eq.~\eqref{eqn:alpha} rarely spans more than one decade in $T$. The range of $V$, where the \textit{non}-Ohmic conductance follows the ``universal" curve in the experiments, is usually less than a decade. In our case, it is wider than one decade. Still, the dependences that we find numerically are not true power-laws. If we look at wider ranges of $V$ and $T$, the deviations are seen. Therefore, our numerical results for the VRH transport, just like the experiments, demonstrate only the apparent power-law behavior (APLB) restricted to a certain parameter range.

\begin{figure}[t]
  \includegraphics[width=2.4in]{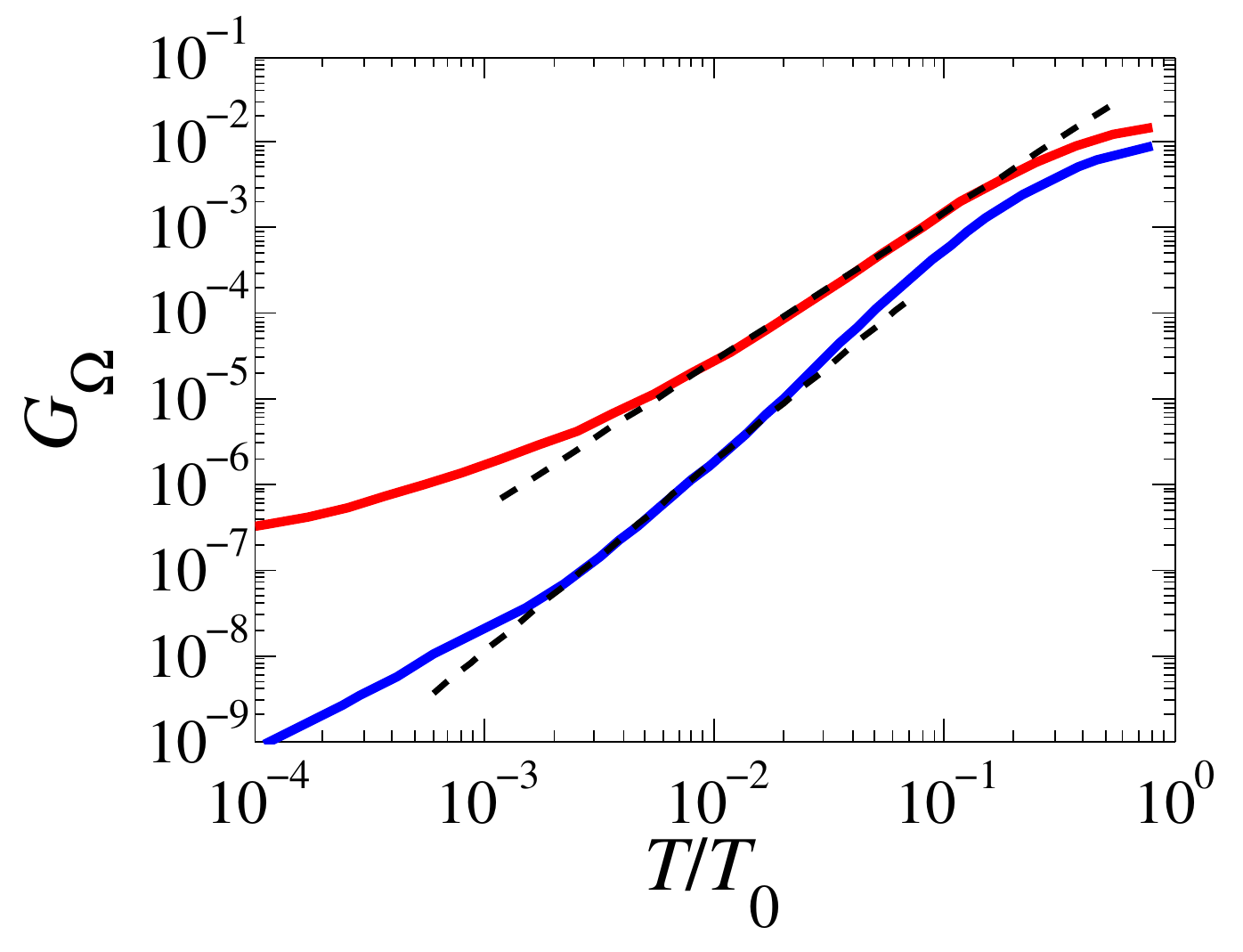}
  \caption{\label{fig:Ohmic_Conductance} (Color online) The Ohmic conductance \textit{vs}. temperature for $L = 30$ (upper curve) and $L = 50$ (lower curve) with $a = 4$. The dashed lines serve to illustrate the apparent linearity of the curves near their inflection points (dots).}
\end{figure}

In our calculations, this range is located near the inflection point of the curve $\ln G_{\Omega}$ \textit{vs}. $\ln T$, see Fig.~\ref{fig:Ohmic_Conductance}. Near the corresponding temperature $T_{\text{inf}}$ the curve can be approximated by a straight line with a certain slope $\alpha$, in agreement with Eq.~\eqref{eqn:alpha}. Further  analysis, following Refs.~\onlinecite{Tartakovskii1987hco, Levin1988thc, Glazman1988ita}, which is discussed below, leads to analytical estimates 
\begin{align}
\alpha &= N_{\text{inf}} - 1 - \frac{2}{N_{\text{inf}}}\,, \quad
N_{\text{inf}} = c_1 \sqrt{\frac{L}{a}}\,,
\label{eqn:alpha_anal}\\
\beta + 1 &= c_2 \alpha\,, \quad
\gamma = c_3\, \frac{2 \pi a}{L}\,,
\label{eqn:gamma}\\
T_{\text{inf}} &= c_4 T_0\, \frac{a}{L}\,, \quad
T_0 \equiv \frac{1}{g a}\,.
\label{eqn:T_inf}
\end{align}
Here $g$ is the density of states and $c_i$'s are coefficients of the order of unity. In comparison, our simulations give $\alpha = 1.75$, $\beta = 1.1$, and $\gamma = 1$ for $L / a = 7.5$. For $L / a = 12.5$, we get $\alpha = 2.4$, $\beta = 1.7$, and $\gamma = 0.6$. This implies $c_1 \approx 1.1$, $c_2 \approx 0.85$, $c_3 \approx 1.2$, and $c_4 \approx 0.4$.

Our numerical results are comparable with typical experimental numbers. They are also consistent with the observed trend that longer and more disordered wires produce larger $\alpha$ and $\beta$ but smaller $\gamma$. A more detailed comparison would require taking into account particularities of a given set of samples beyond our generic model. Due to individual variations in the nature of disorder and the parameters of electron-phonon coupling, $\alpha$ and $\beta$ may acquire additional corrections of the order of unity. 


%
%

\begin{figure}[b]
  \includegraphics[width=3.5in]{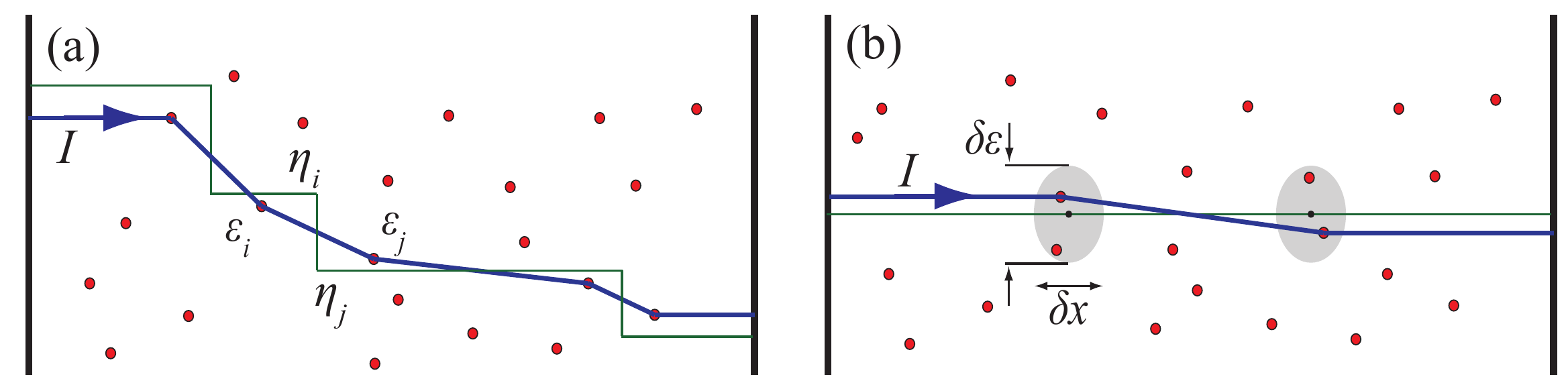}
  \caption{\label{fig:Paths} (a) A typical hopping path through the wire (thick line). The thin line represents electrochemical potential $\eta$. (b) A rare path \cite{Tartakovskii1987hco, Levin1988thc, Glazman1988ita,Bahlouli1994cci} made of equal-length hops. Here $\delta\epsilon \sim NT$ and $\delta x \sim Na$, where $N$ is the number of hops. Weakly varying $\eta(x)$ corresponds to the Ohmic regime.}
\end{figure}

Let us now give more details regarding the calculations. We consider a system of localized states with random energies $\varepsilon_i$ (Fig.~\ref{fig:Paths}) distributed according to the Poisson distribution with the average energy separation $\Delta = 3$. We treat all $\varepsilon_i$ as constants, independent of the applied current. This is justified if electron interactions are weak~\cite{Rodin2009nso}. The $x$-spacing between the sites is taken to be unity, so that the density of states is $g = 1 / \Delta = 1 / 3$. The localization length $a$ is chosen to be $4$. To avoid repeated negative signs in the formulas, we take the electron charge $e = 1$ to be positive.

Let $\eta_i$ be the electrochemical potential of site $i$, then the net current from site $i$ to another site $j$ is given by~\cite{Pollak1976apt}
\begin{equation}
I_{i j} = \frac{\Gamma_{i j} \exp\left(-\frac{2}{a} x_{i j}\right) \sinh\left(\frac{\eta_i - \eta_j}{2T}\right)}{\cosh\left(\frac{\varepsilon_i - \eta_i}{2T}\right)\cosh\left(\frac{\varepsilon_j - \eta_j}{2T}\right)\sinh\left|\frac{\varepsilon_i - \varepsilon_j}{2T}\right|}\,,
\label{eqn:Current}
\end{equation}
where $x_{i j} \equiv |x_i - x_j|$, $\Gamma_{i j} = G_0\, |\varepsilon_i - \varepsilon_j|$, and parameter $G_0$ of dimension of conductance is related to electron-phonon coupling~\cite{Miller1960ica,  Shklovskii1984epo, Glazman1988ita}. The problem is to determine $I_{i j}$ and $\eta_i$ that satisfy the current conservation.

The analysis is simplified by the conventional assumption that the transport is dominated by a single optimal path of least resistance. Within this approximation, the current does not branch, i.e., $I_{i j} = I$ in each link of the path. The total voltage drop $V$ across the sample is the sum of voltage drops $\eta_{i} - \eta_{j}$ on the links. One can determine the optimal path by finding the sequence of sites that gives the smallest $V$ for a given $I$~\cite{Rodin2009nso}. 

Let us define auxiliary variables $u_I \equiv \ln(T G_0 / I)$ and 
\begin{equation}
S \equiv \frac{T G_0}{\Gamma_{i j}} \exp\left(\frac{2x_{i j}}{a}\right) \cosh\, \frac{\varepsilon_i - \eta_i}{2T}\, \sinh \, \frac{|\varepsilon_i - \varepsilon_j|}{2 T}\,.
\label{eqn:S}
\end{equation}
Solving Eq.~\eqref{eqn:Current} for $\eta_j$, we obtain
\begin{equation}
\eta_j = T \ln \left( \frac{e^{\eta_i / 2T} - S e^{-u_I + \varepsilon_j / 2 T}} {e^{-\eta_i / 2T} + S e^{-u_I -\varepsilon_j / 2 T}}\right).
\label{eqn:etaj}
\end{equation}
Unlike the internal hops, transitions between the source electrode and the first site inside the sample (as well as the last site and the drain) do not require phonons. This can be accounted for by using
\begin{equation}
S_c = \frac{2 G_0}{G_c} \exp \left(\frac{2 x_{i j}}{a}\right) \cosh \left(\frac{\varepsilon - \eta_i}{2 T}\right)
\label{eqn:S_c}
\end{equation}
in lieu of $S$. Here $G_c$ is determined by the tunneling transparency of the contact between the sample and the electrode. We choose a representative value $G_c = 4 G_0$. Note that the numerator of Eq.~\eqref{eqn:etaj} must be positive, which sets a limit on the maximum current that can flow between sites $i$ and $j$.

To find the optimal path, we use a modified Dijkstra algorithm~\cite{Rodin2009nso}, in which the ``cost" of reaching site $j$ starting from the source equals $-\eta_j$. The latter is calculated recurrently using Eq.~\eqref{eqn:Current}. For each disorder realization the resistance $V / I$ is random and by running the simulations many times we can compute its probability distribution. Taking the average over the latter as explained in Ref.~\cite{Rodin2009nso} we get the ensemble-averaged $G(V, T)$.

The results for the Ohmic regime (Fig.~\ref{fig:Ohmic_Conductance}) were obtained by choosing a very large $u_I = 40$ to ensure $V \ll T$. We analyzed two different system lengths: $L = 30 = 7.5 a$ and $L = 50 = 12.5 a$. For $L = 30$ we generated an ensemble of $20,000$ samples and for $L = 50$ we used $10,000$ samples in order to average out the mesoscopic fluctuations. (Actually, using 500 samples would give results of comparable quality.) Figure~\ref{fig:Ohmic_Conductance} clearly demonstrates more than a decade of the APLB of Eq.~\eqref{eqn:alpha} near the inflection points of the curves. Note that this point is located at a lower temperature for the longer sample.

Having determined the range of $T$ where we get the Ohmic APLB, we proceeded to analyzing the non-Ohmic behavior of the system in this range of temperatures. To this end we fitted the results for higher $V$ to Eq.~\eqref{eqn:LL}. For $L = 30$, Fig.~\ref{fig:L30_Collapse}, we found a good collapse in both the Ohmic and non-Ohmic regimes. All curves in Fig.~\ref{fig:L30_Collapse} were cut at $V = 2$, since at that point the curves were beginning to saturate as they approached the maximum current possible in the system. The collapse obtained for $L = 50$ (not shown) was equally good. The quality of our data collapse matches or exceeds that in the experiments~\cite{Aleshin2004odt, Slot2004odc, Venkataraman2006eti, Zhou2007ode, Coiffic2007cri, Rahman2010bdc, Yuen2009nti, Worne2010tio}. The values of the fitting parameters $\alpha$, $\beta$, and $\gamma$ have already been discussed (see more below).

%
%

%
%

Let us now examine how the APLB we have found numerically can be understood in the light of established theory of 1D VRH. According to this theory, transport is characterized by several regimes. At low $T$, the conductance of the ensemble is dominated by rare paths with nearly equidistant sites, see Fig.~\ref{fig:Paths}(b). This regime was studied in Refs.~\onlinecite{Glazman1988ita, Bahlouli1994cci, Ballard2006roc} for two intermediate sites and in Refs.~\onlinecite{Tartakovskii1987hco, Levin1988thc} for a chain of many sites~\cite{[{A near-literal copy of Ref.~\protect{\onlinecite{Tartakovskii1987hco}} is available online as }] Park2000thc}. Adopting the derivation in Ref.~\onlinecite{Tartakovskii1987hco} to the 1D case we can show that at a given $T$ the main contribution to $G_{\Omega}$ comes from the chains of $N = \sqrt{2 L / \lambda a}$ hops, where $\lambda$ is the solution of the equation $\lambda \simeq \ln(\lambda / L g T)$. In addition,~\cite{Glazman1988ita, Bahlouli1994cci, Xu1995dih}
\begin{equation}
\frac{d \ln G_{\Omega}}{d \ln T} \simeq N - 1 - \frac{2}{N}\,.
\label{eqn:dG_dT}
\end{equation}
At $T = T_{\text{inf}}$, we have $\lambda \sim 1$, which yields Eq.~\eqref{eqn:alpha_anal}.

For $T > T_{\text{inf}}$, the system enters the regime where the transport is limited by rare highly resistive links --- ``breaks'' --- on the optimal path~\cite{Kurkijarvi1973hci, Lee1984vrh, Raikh1989fot}. As a result, the Ohmic conductance, which can be derived from the formulas of Ref.~\onlinecite{Raikh1989fot}, obeys Eq.~\eqref{eqn:VRH} with the coefficients
\begin{equation}
\mu = \frac12\,, \quad
T_* \simeq 2 T_0 \ln\left(\frac{2 L e}{a}\, \frac{T}{T_0}\right)\,.
\label{eqn:RR_G}
\end{equation}
The concavity of the $\ln G_{\Omega}$ \textit{vs}. $\ln T$ curve is opposite in the two temperature ranges, which creates the inflection point, see Fig.~\ref{fig:Ohmic_Conductance}.

The non-Ohmic transport is also characterized by an $S$-shaped curve of $\ln I$ \textit{vs}. $\ln V$, with its own inflection point. For example, at $T \gg T_{\text{inf}}$, the theory~\cite{Fogler2005nov} predicts
\begin{equation}
\ln \frac{V}{T_0} = -\frac{u_I^2}{u_M^2}+\ln\left(\frac{8L/a}{u_I^2}\right)\,, \quad u_M \equiv \sqrt{\frac{2T_0}{T}}\,.
\label{eqn:FK}
\end{equation}
By the argument similar to that used in the Ohmic regime, $\beta + 1$ in Eq.~\eqref{eqn:beta} is determined by the maximum slope, i.e., the derivative of $\ln I$ with respect to $\ln V$:
\begin{equation}
\beta + 1 = \max \left(-\frac{d u_I}{d \ln V}\right) = \frac{u_M}{4} = \sqrt{\frac{T_0}{8T}}\,.
\label{eqn:betaFK}
\end{equation}
We see that $\beta + 1 \propto T^{-1/2}$ is not a constant but decreases with $T$, in qualitative agreement with experiments~\cite{Zhou2007ode, Rahman2010bdc, Coiffic2007cri}. This explains why the data collapse onto the universal curve of Eq.~\eqref{eqn:LL} can be achieved only in a limited range of $T$.

At $T = T_\text{inf}$, Eq.~\eqref{eqn:FK} is at the border of its validity. Hence, Eq.~\eqref{eqn:betaFK} gives only the order of magnitude estimate, $\beta + 1 \lesssim \alpha$, which is the first part of Eq.~\eqref{eqn:gamma}. Finally, to get $\gamma$ we note that the crossover to the $T$-independent behavior in Eq.~\eqref{eqn:LL} takes place at $\gamma V \sim 2\pi T$. On the other hand, according to Eq.~\eqref{eqn:FK}, this occurs at $u_I \sim u_M$ where $V / T_0 \sim L T / a T_0$. Combining these expressions, we recover the second part  of Eq.~\eqref{eqn:gamma}.

Formulas~\eqref{eqn:alpha_anal}--\eqref{eqn:gamma} predict numerical values and relations among $\alpha$, $\beta$, and $\gamma$ that are in agreement with most of the cited experiments~\cite{Venkataraman2006eti, Slot2004odc, Aleshin2004odt, Zhou2007ode, Rahman2010bdc, Coiffic2007cri, Yuen2009nti, Worne2010tio}. Additionally, they provide a way of estimating the localization length $a$. For example, taking parameters $\alpha = 4.3$, $\beta = 2.1$, $\gamma = 0.25$, $L \sim 1\, \mu\,\text{m}$ of a representative $\text{MoSe}_{2}$ nanowire from Ref.~\onlinecite{Venkataraman2006eti}, we find $a \sim 40\,\text{nm}$ for this sample (W3). Other samples measured in that work showed $\alpha \propto 1 / \sqrt{M}$ scaling with the number of transport channels $M$. In our model the same scaling occurs if $a \propto M$, as in a weakly disordered quasi-1D metal. In such a system $a$ can be enlarged by applying an external magnetic field~\cite{Beenakker1997rmt}. This is one convenient way to further test our model experimentally. Alternatively, it may be possible to vary the disorder strength and therefore $a$ by electrostatic gating, while monitoring the predicted trends in $\alpha$, $\beta$, and $\gamma$.


Another model we considered in search for the APLB was the interrupted-strand model (ISM)~\cite{Fogler2004vrh, Fogler2006cba}. Therein a metallic wire is divided into segments by randomly positioned impurities of tunneling transparency $e^{-s} \ll 1$, which turn it into a chain of weakly coupled quantum dots. In the simulations we studied wires with $N_i = 50$ impurities of strength $s = 4$. While we did observe the APLB in such wires ($\alpha = 3.75$, $\beta = 1.6$, $\gamma = 0.15$), the quality of the data collapse was not as good as in Fig.~\ref{fig:L30_Collapse}. Otherwise, the results were similar~\cite{[{See also }][{, for simulations of longer wires and larger $s$}]Deroulers2007dot}.

Note that the VRH in the ISM is analogous to multi-dot cotunneling in a granular metal. The latter also leads, in all spatial dimensions, to the power-law conductance behavior with $\alpha = \beta = 2 N_d - 4$, $N_d$ being the number of dots involved in one cotunneling event~\cite{Tran2005mci}. Hence, the APLB is not uncommon in the VRH regime.

The authors are grateful to K. Matveev and M. Bockrath for discussion of the results and to M. Di Ventra for providing access to computer resources on which a part of the calculations was carried out. This work is supported by the Grant NSF DMR-0706654.


\begin{thebibliography}{44}%
\makeatletter
\providecommand \@ifxundefined [1]{%
 \@ifx{#1\undefined}
}%
\providecommand \@ifnum [1]{%
 \ifnum #1\expandafter \@firstoftwo
 \else \expandafter \@secondoftwo
 \fi
}%
\providecommand \@ifx [1]{%
 \ifx #1\expandafter \@firstoftwo
 \else \expandafter \@secondoftwo
 \fi
}%
\providecommand \natexlab [1]{#1}%
\providecommand \enquote  [1]{``#1''}%
\providecommand \bibnamefont  [1]{#1}%
\providecommand \bibfnamefont [1]{#1}%
\providecommand \citenamefont [1]{#1}%
\providecommand \href@noop [0]{\@secondoftwo}%
\providecommand \href [0]{\begingroup \@sanitize@url \@href}%
\providecommand \@href[1]{\@@startlink{#1}\@@href}%
\providecommand \@@href[1]{\endgroup#1\@@endlink}%
\providecommand \@sanitize@url [0]{\catcode `\\12\catcode `\$12\catcode
  `\&12\catcode `\#12\catcode `\^12\catcode `\_12\catcode `\%12\relax}%
\providecommand \@@startlink[1]{}%
\providecommand \@@endlink[0]{}%
\providecommand \url  [0]{\begingroup\@sanitize@url \@url }%
\providecommand \@url [1]{\endgroup\@href {#1}{\urlprefix }}%
\providecommand \urlprefix  [0]{URL }%
\providecommand \Eprint [0]{\href }%
\@ifxundefined \urlstyle {%
  \providecommand \doi  [0]{\begingroup \@sanitize@url \@doi}%
  \providecommand \@doi [1]{\endgroup \@@startlink {\doibase
  #1}doi:\discretionary {}{}{}#1\@@endlink }%
}{%
  \providecommand \doi  [0]{doi:\discretionary{}{}{}\begingroup
  \urlstyle{rm}\Url }%
}%
\providecommand \doibase [0]{http://dx.doi.org/}%
\providecommand \Doi [0]{\begingroup \@sanitize@url \@Doi }%
\providecommand \@Doi  [1]{\endgroup\@@startlink{\doibase#1}\@@Doi}%
\providecommand \@@Doi [1]{#1\@@endlink}%
\providecommand \selectlanguage [0]{\@gobble}%
\providecommand \bibinfo  [0]{\@secondoftwo}%
\providecommand \bibfield  [0]{\@secondoftwo}%
\providecommand \translation [1]{[#1]}%
\providecommand \BibitemOpen [0]{}%
\providecommand \bibitemStop [0]{}%
\providecommand \bibitemNoStop [0]{.\EOS\space}%
\providecommand \EOS [0]{\spacefactor3000\relax}%
\providecommand \BibitemShut  [1]{\csname bibitem#1\endcsname}%
\bibitem [{\citenamefont {Shklovskii}\ and\ \citenamefont
  {Efros}(1984)}]{Shklovskii1984epo}%
  \BibitemOpen
  \bibfield  {author} {\bibinfo {author} {\bibfnamefont {B.~I.}\ \bibnamefont
  {Shklovskii}}\ and\ \bibinfo {author} {\bibfnamefont {A.~L.}\ \bibnamefont
  {Efros}},\ }\href@noop {} {\emph {\bibinfo {title} {Electronic Properties of
  Doped Semiconductors}}}\ (\bibinfo  {publisher} {Springer-Verlag},\ \bibinfo
  {address} {Berlin},\ \bibinfo {year} {1984})\BibitemShut {NoStop}%
\bibitem [{\citenamefont {Bockrath}\ \emph {et~al.}(1999)\citenamefont
  {Bockrath}, \citenamefont {Cobden}, \citenamefont {Lu}, \citenamefont
  {Rinzler}, \citenamefont {Smalley}, \citenamefont {Balents},\ and\
  \citenamefont {McEuen}}]{Bockrath1999llb}%
  \BibitemOpen
  \bibfield  {author} {\bibinfo {author} {\bibfnamefont {M.}~\bibnamefont
  {Bockrath}}, \bibinfo {author} {\bibfnamefont {D.~H.}\ \bibnamefont
  {Cobden}}, \bibinfo {author} {\bibfnamefont {J.}~\bibnamefont {Lu}}, \bibinfo
  {author} {\bibfnamefont {A.~G.}\ \bibnamefont {Rinzler}}, \bibinfo {author}
  {\bibfnamefont {R.~E.}\ \bibnamefont {Smalley}}, \bibinfo {author}
  {\bibfnamefont {L.}~\bibnamefont {Balents}}, \ and\ \bibinfo {author}
  {\bibfnamefont {P.~L.}\ \bibnamefont {McEuen}},\ }\Doi {10.1038/17569}
  {\bibfield  {journal} {\bibinfo  {journal} {Nature},\ }\textbf {\bibinfo
  {volume} {397}},\ \bibinfo {pages} {598} (\bibinfo {year}
  {1999})}\BibitemShut {NoStop}%
\bibitem [{\citenamefont {Yao}\ \emph {et~al.}(1999)\citenamefont {Yao},
  \citenamefont {Postma}, \citenamefont {Balents},\ and\ \citenamefont
  {Dekker}}]{Yao1999cni}%
  \BibitemOpen
  \bibfield  {author} {\bibinfo {author} {\bibfnamefont {Z.}~\bibnamefont
  {Yao}}, \bibinfo {author} {\bibfnamefont {H.~W.~C.}\ \bibnamefont {Postma}},
  \bibinfo {author} {\bibfnamefont {L.}~\bibnamefont {Balents}}, \ and\
  \bibinfo {author} {\bibfnamefont {C.}~\bibnamefont {Dekker}},\ }\Doi
  {10.1038/46241} {\bibfield  {journal} {\bibinfo  {journal} {Nature},\
  }\textbf {\bibinfo {volume} {402}},\ \bibinfo {pages} {273} (\bibinfo {year}
  {1999})}\BibitemShut {NoStop}%
\bibitem [{\citenamefont {Bachtold}\ \emph {et~al.}(2001)\citenamefont
  {Bachtold}, \citenamefont {de~Jonge}, \citenamefont {Grove-Rasmussen},
  \citenamefont {McEuen}, \citenamefont {Buitelaar},\ and\ \citenamefont
  {Sch\"onenberger}}]{Bachtold2001sot}%
  \BibitemOpen
  \bibfield  {author} {\bibinfo {author} {\bibfnamefont {A.}~\bibnamefont
  {Bachtold}}, \bibinfo {author} {\bibfnamefont {M.}~\bibnamefont {de~Jonge}},
  \bibinfo {author} {\bibfnamefont {K.}~\bibnamefont {Grove-Rasmussen}},
  \bibinfo {author} {\bibfnamefont {P.~L.}\ \bibnamefont {McEuen}}, \bibinfo
  {author} {\bibfnamefont {M.}~\bibnamefont {Buitelaar}}, \ and\ \bibinfo
  {author} {\bibfnamefont {C.}~\bibnamefont {Sch\"onenberger}},\ }\Doi
  {10.1103/PhysRevLett.87.166801} {\bibfield  {journal} {\bibinfo  {journal}
  {Phys. Rev. Lett.},\ }\textbf {\bibinfo {volume} {87}},\ \bibinfo {pages}
  {166801} (\bibinfo {year} {2001})}\BibitemShut {NoStop}%
\bibitem [{\citenamefont {Kanda}\ \emph {et~al.}(2004)\citenamefont {Kanda},
  \citenamefont {Tsukagoshi}, \citenamefont {Aoyagi},\ and\ \citenamefont
  {Ootuka}}]{Kanda2004gvd}%
  \BibitemOpen
  \bibfield  {author} {\bibinfo {author} {\bibfnamefont {A.}~\bibnamefont
  {Kanda}}, \bibinfo {author} {\bibfnamefont {K.}~\bibnamefont {Tsukagoshi}},
  \bibinfo {author} {\bibfnamefont {Y.}~\bibnamefont {Aoyagi}}, \ and\ \bibinfo
  {author} {\bibfnamefont {Y.}~\bibnamefont {Ootuka}},\ }\Doi
  {10.1103/PhysRevLett.92.036801} {\bibfield  {journal} {\bibinfo  {journal}
  {Phys. Rev. Lett.},\ }\textbf {\bibinfo {volume} {92}},\ \bibinfo {pages}
  {036801} (\bibinfo {year} {2004})}\BibitemShut {NoStop}%
\bibitem [{\citenamefont {Gao}\ \emph {et~al.}(2004)\citenamefont {Gao},
  \citenamefont {Komnik}, \citenamefont {Egger}, \citenamefont {Glattli},\ and\
  \citenamefont {Bachtold}}]{Gao2004efl}%
  \BibitemOpen
  \bibfield  {author} {\bibinfo {author} {\bibfnamefont {B.}~\bibnamefont
  {Gao}}, \bibinfo {author} {\bibfnamefont {A.}~\bibnamefont {Komnik}},
  \bibinfo {author} {\bibfnamefont {R.}~\bibnamefont {Egger}}, \bibinfo
  {author} {\bibfnamefont {D.~C.}\ \bibnamefont {Glattli}}, \ and\ \bibinfo
  {author} {\bibfnamefont {A.}~\bibnamefont {Bachtold}},\ }\Doi
  {10.1103/PhysRevLett.92.216804} {\bibfield  {journal} {\bibinfo  {journal}
  {Phys. Rev. Lett.},\ }\textbf {\bibinfo {volume} {92}},\ \bibinfo {pages}
  {216804} (\bibinfo {year} {2004})}\BibitemShut {NoStop}%
\bibitem [{\citenamefont {Monteverde}\ \emph {et~al.}(2006)\citenamefont
  {Monteverde}, \citenamefont {Garbarino}, \citenamefont
  {N{\'u}{\~n}ez-Reguiero}, \citenamefont {Souletie}, \citenamefont {Acha},
  \citenamefont {Jing}, \citenamefont {Lu}, \citenamefont {Pan}, \citenamefont
  {Xie},\ and\ \citenamefont {Egger}}]{Monteverde2006tll}%
  \BibitemOpen
  \bibfield  {author} {\bibinfo {author} {\bibfnamefont {M.}~\bibnamefont
  {Monteverde}}, \bibinfo {author} {\bibfnamefont {G.}~\bibnamefont
  {Garbarino}}, \bibinfo {author} {\bibfnamefont {M.}~\bibnamefont
  {N{\'u}{\~n}ez-Reguiero}}, \bibinfo {author} {\bibfnamefont {J.}~\bibnamefont
  {Souletie}}, \bibinfo {author} {\bibfnamefont {C.}~\bibnamefont {Acha}},
  \bibinfo {author} {\bibfnamefont {X.}~\bibnamefont {Jing}}, \bibinfo {author}
  {\bibfnamefont {L.}~\bibnamefont {Lu}}, \bibinfo {author} {\bibfnamefont
  {Z.~W.}\ \bibnamefont {Pan}}, \bibinfo {author} {\bibfnamefont {S.~S.}\
  \bibnamefont {Xie}}, \ and\ \bibinfo {author} {\bibfnamefont
  {R.}~\bibnamefont {Egger}},\ }\Doi {10.1103/PhysRevLett.97.176401} {\bibfield
   {journal} {\bibinfo  {journal} {Phys. Rev. Lett.},\ }\textbf {\bibinfo
  {volume} {97}},\ \bibinfo {pages} {176401} (\bibinfo {year}
  {2006})}\BibitemShut {NoStop}%
\bibitem [{\citenamefont {Coiffic}\ \emph {et~al.}(2007)\citenamefont
  {Coiffic}, \citenamefont {Fayolle}, \citenamefont {Maitrejean}, \citenamefont
  {{Foa~Torres}},\ and\ \citenamefont {Poche}}]{Coiffic2007cri}%
  \BibitemOpen
  \bibfield  {author} {\bibinfo {author} {\bibfnamefont {J.~C.}\ \bibnamefont
  {Coiffic}}, \bibinfo {author} {\bibfnamefont {M.}~\bibnamefont {Fayolle}},
  \bibinfo {author} {\bibfnamefont {S.}~\bibnamefont {Maitrejean}}, \bibinfo
  {author} {\bibfnamefont {L.~E.~F.}\ \bibnamefont {{Foa~Torres}}}, \ and\
  \bibinfo {author} {\bibfnamefont {H.~L.}\ \bibnamefont {Poche}},\ }\Doi
  {10.1063/1.2826274} {\bibfield  {journal} {\bibinfo  {journal} {Appl. Phys.
  Lett.},\ }\textbf {\bibinfo {volume} {91}},\ \bibinfo {pages} {252107}
  (\bibinfo {year} {2007})}\BibitemShut {NoStop}%
\bibitem [{\citenamefont {Dayen}\ \emph {et~al.}(2009)\citenamefont {Dayen},
  \citenamefont {T.L.Wade}, \citenamefont {Rizza}, \citenamefont {Golubev},
  \citenamefont {Cojocaru}, \citenamefont {Pribat}, \citenamefont {Jehl},
  \citenamefont {Sanquer},\ and\ \citenamefont {J.-E.Wegrowe}}]{Dayen2009cod}%
  \BibitemOpen
  \bibfield  {author} {\bibinfo {author} {\bibfnamefont {J.-F.}\ \bibnamefont
  {Dayen}}, \bibinfo {author} {\bibnamefont {T.L.Wade}}, \bibinfo {author}
  {\bibfnamefont {G.}~\bibnamefont {Rizza}}, \bibinfo {author} {\bibfnamefont
  {D.}~\bibnamefont {Golubev}}, \bibinfo {author} {\bibfnamefont {C.-S.}\
  \bibnamefont {Cojocaru}}, \bibinfo {author} {\bibfnamefont {D.}~\bibnamefont
  {Pribat}}, \bibinfo {author} {\bibfnamefont {X.}~\bibnamefont {Jehl}},
  \bibinfo {author} {\bibfnamefont {M.}~\bibnamefont {Sanquer}}, \ and\
  \bibinfo {author} {\bibnamefont {J.-E.Wegrowe}},\ }\Doi
  {10.1051/epjap/2009132} {\bibfield  {journal} {\bibinfo  {journal} {Eur.
  Phys. J. Appl. Phys.},\ }\textbf {\bibinfo {volume} {48}},\ \bibinfo {pages}
  {10604} (\bibinfo {year} {2009})}\BibitemShut {NoStop}%
\bibitem [{\citenamefont {Zaitsev-Zotov}\ \emph {et~al.}(2000)\citenamefont
  {Zaitsev-Zotov}, \citenamefont {Kumzerov}, \citenamefont {Firsov},\ and\
  \citenamefont {Monceau}}]{Zaitsev2000lll}%
  \BibitemOpen
  \bibfield  {author} {\bibinfo {author} {\bibfnamefont {S.~V.}\ \bibnamefont
  {Zaitsev-Zotov}}, \bibinfo {author} {\bibfnamefont {Y.~A.}\ \bibnamefont
  {Kumzerov}}, \bibinfo {author} {\bibfnamefont {Y.~A.}\ \bibnamefont
  {Firsov}}, \ and\ \bibinfo {author} {\bibfnamefont {P.}~\bibnamefont
  {Monceau}},\ }\Doi {10.1088/0953-8984/12/20/101} {\bibfield  {journal}
  {\bibinfo  {journal} {J.Phys.:Condens. Matter},\ }\textbf {\bibinfo {volume}
  {12}},\ \bibinfo {pages} {303} (\bibinfo {year} {2000})}\BibitemShut
  {NoStop}%
\bibitem [{\citenamefont {Chang}(2003)}]{Chang2003cll}%
  \BibitemOpen
  \bibfield  {author} {\bibinfo {author} {\bibfnamefont {A.~M.}\ \bibnamefont
  {Chang}},\ }\Doi {10.1103/RevModPhys.75.1449} {\bibfield  {journal} {\bibinfo
   {journal} {Rev. Mod. Phys.},\ }\textbf {\bibinfo {volume} {75}},\ \bibinfo
  {pages} {1449} (\bibinfo {year} {2003})}\BibitemShut {NoStop}%
\bibitem [{\citenamefont {Tserkovnyak}\ \emph {et~al.}(2003)\citenamefont
  {Tserkovnyak}, \citenamefont {Halperin}, \citenamefont {Auslaender},\ and\
  \citenamefont {Yacoby}}]{Tserkovnyak2003iaz}%
  \BibitemOpen
  \bibfield  {author} {\bibinfo {author} {\bibfnamefont {Y.}~\bibnamefont
  {Tserkovnyak}}, \bibinfo {author} {\bibfnamefont {B.~I.}\ \bibnamefont
  {Halperin}}, \bibinfo {author} {\bibfnamefont {O.~M.}\ \bibnamefont
  {Auslaender}}, \ and\ \bibinfo {author} {\bibfnamefont {A.}~\bibnamefont
  {Yacoby}},\ }\Doi {10.1103/PhysRevB.68.125312} {\bibfield  {journal}
  {\bibinfo  {journal} {Phys. Rev. B},\ }\textbf {\bibinfo {volume} {68}},\
  \bibinfo {pages} {125312} (\bibinfo {year} {2003})}\BibitemShut {NoStop}%
\bibitem [{\citenamefont {Slot}\ \emph {et~al.}(2004)\citenamefont {Slot},
  \citenamefont {Holst}, \citenamefont {van~der Zant},\ and\ \citenamefont
  {Zaitsev-Zotov}}]{Slot2004odc}%
  \BibitemOpen
  \bibfield  {author} {\bibinfo {author} {\bibfnamefont {E.}~\bibnamefont
  {Slot}}, \bibinfo {author} {\bibfnamefont {M.~A.}\ \bibnamefont {Holst}},
  \bibinfo {author} {\bibfnamefont {H.~S.~J.}\ \bibnamefont {van~der Zant}}, \
  and\ \bibinfo {author} {\bibfnamefont {S.~V.}\ \bibnamefont
  {Zaitsev-Zotov}},\ }\Doi {10.1103/PhysRevLett.93.176602} {\bibfield
  {journal} {\bibinfo  {journal} {Phys. Rev. Lett.},\ }\textbf {\bibinfo
  {volume} {93}},\ \bibinfo {pages} {176602} (\bibinfo {year}
  {2004})}\BibitemShut {NoStop}%
\bibitem [{\citenamefont {Aleshin}\ \emph {et~al.}(2004)\citenamefont
  {Aleshin}, \citenamefont {Lee}, \citenamefont {Park},\ and\ \citenamefont
  {Akagi}}]{Aleshin2004odt}%
  \BibitemOpen
  \bibfield  {author} {\bibinfo {author} {\bibfnamefont {A.~N.}\ \bibnamefont
  {Aleshin}}, \bibinfo {author} {\bibfnamefont {H.~J.}\ \bibnamefont {Lee}},
  \bibinfo {author} {\bibfnamefont {Y.~W.}\ \bibnamefont {Park}}, \ and\
  \bibinfo {author} {\bibfnamefont {K.}~\bibnamefont {Akagi}},\ }\Doi
  {10.1103/PhysRevLett.93.196601} {\bibfield  {journal} {\bibinfo  {journal}
  {Phys. Rev. Lett.},\ }\textbf {\bibinfo {volume} {93}},\ \bibinfo {pages}
  {196601} (\bibinfo {year} {2004})}\BibitemShut {NoStop}%
\bibitem [{\citenamefont {Rahman}\ and\ \citenamefont
  {Sanyal}(2010)}]{Rahman2010bdc}%
  \BibitemOpen
  \bibfield  {author} {\bibinfo {author} {\bibfnamefont {A.}~\bibnamefont
  {Rahman}}\ and\ \bibinfo {author} {\bibfnamefont {M.~K.}\ \bibnamefont
  {Sanyal}},\ }\Doi {10.1088/0953-8984/22/17/175301} {\bibfield  {journal}
  {\bibinfo  {journal} {J. Phys.: Condens. Matter},\ }\textbf {\bibinfo
  {volume} {22}},\ \bibinfo {pages} {175301} (\bibinfo {year}
  {2010})}\BibitemShut {NoStop}%
\bibitem [{\citenamefont {Venkataraman}\ \emph {et~al.}(2006)\citenamefont
  {Venkataraman}, \citenamefont {Hong},\ and\ \citenamefont
  {Kim}}]{Venkataraman2006eti}%
  \BibitemOpen
  \bibfield  {author} {\bibinfo {author} {\bibfnamefont {L.}~\bibnamefont
  {Venkataraman}}, \bibinfo {author} {\bibfnamefont {Y.~S.}\ \bibnamefont
  {Hong}}, \ and\ \bibinfo {author} {\bibfnamefont {P.}~\bibnamefont {Kim}},\
  }\Doi {10.1103/PhysRevLett.96.076601} {\bibfield  {journal} {\bibinfo
  {journal} {Phys. Rev. Lett.},\ }\textbf {\bibinfo {volume} {96}},\ \bibinfo
  {pages} {076601} (\bibinfo {year} {2006})}\BibitemShut {NoStop}%
\bibitem [{\citenamefont {Zhou}\ \emph {et~al.}(2007)\citenamefont {Zhou},
  \citenamefont {Xiao}, \citenamefont {Jin}, \citenamefont {Mandrus},
  \citenamefont {Tao}, \citenamefont {Geohegan},\ and\ \citenamefont
  {Pennycook}}]{Zhou2007ode}%
  \BibitemOpen
  \bibfield  {author} {\bibinfo {author} {\bibfnamefont {Z.}~\bibnamefont
  {Zhou}}, \bibinfo {author} {\bibfnamefont {K.}~\bibnamefont {Xiao}}, \bibinfo
  {author} {\bibfnamefont {R.}~\bibnamefont {Jin}}, \bibinfo {author}
  {\bibfnamefont {D.}~\bibnamefont {Mandrus}}, \bibinfo {author} {\bibfnamefont
  {J.}~\bibnamefont {Tao}}, \bibinfo {author} {\bibfnamefont {D.}~\bibnamefont
  {Geohegan}}, \ and\ \bibinfo {author} {\bibfnamefont {S.}~\bibnamefont
  {Pennycook}},\ }\Doi {10.1063/1.2738380} {\bibfield  {journal} {\bibinfo
  {journal} {Appl. Phys. Lett.},\ }\textbf {\bibinfo {volume} {90}},\ \bibinfo
  {pages} {193115} (\bibinfo {year} {2007})}\BibitemShut {NoStop}%
\bibitem [{\citenamefont {Yuen}\ \emph {et~al.}(2009)\citenamefont {Yuen},
  \citenamefont {Menon}, \citenamefont {Coates}, \citenamefont {Namdas},
  \citenamefont {Cho}, \citenamefont {Hannahs}, \citenamefont {Moses},\ and\
  \citenamefont {Heeger}}]{Yuen2009nti}%
  \BibitemOpen
  \bibfield  {author} {\bibinfo {author} {\bibfnamefont {J.~D.}\ \bibnamefont
  {Yuen}}, \bibinfo {author} {\bibfnamefont {R.}~\bibnamefont {Menon}},
  \bibinfo {author} {\bibfnamefont {N.~E.}\ \bibnamefont {Coates}}, \bibinfo
  {author} {\bibfnamefont {E.~B.}\ \bibnamefont {Namdas}}, \bibinfo {author}
  {\bibfnamefont {S.}~\bibnamefont {Cho}}, \bibinfo {author} {\bibfnamefont
  {S.~T.}\ \bibnamefont {Hannahs}}, \bibinfo {author} {\bibfnamefont
  {D.}~\bibnamefont {Moses}}, \ and\ \bibinfo {author} {\bibfnamefont {A.~J.}\
  \bibnamefont {Heeger}},\ }\Doi {10.1038/nmat2470} {\bibfield  {journal}
  {\bibinfo  {journal} {Nature Mater.},\ }\textbf {\bibinfo {volume} {8}},\
  \bibinfo {pages} {572} (\bibinfo {year} {2009})}\BibitemShut {NoStop}%
\bibitem [{\citenamefont {Worne}\ \emph {et~al.}(2010)\citenamefont {Worne},
  \citenamefont {Anthony},\ and\ \citenamefont {Natelson}}]{Worne2010tio}%
  \BibitemOpen
  \bibfield  {author} {\bibinfo {author} {\bibfnamefont {J.~H.}\ \bibnamefont
  {Worne}}, \bibinfo {author} {\bibfnamefont {J.~E.}\ \bibnamefont {Anthony}},
  \ and\ \bibinfo {author} {\bibfnamefont {D.}~\bibnamefont {Natelson}},\ }\Doi
  {10.1063/1.3309704} {\bibfield  {journal} {\bibinfo  {journal} {Appl. Phys.
  Lett.},\ }\textbf {\bibinfo {volume} {96}},\ \bibinfo {pages} {053308}
  (\bibinfo {year} {2010})}\BibitemShut {NoStop}%
\bibitem [{\citenamefont {Giamarchi}(2004)}]{Giamarchi2004qpi}%
  \BibitemOpen
  \bibfield  {author} {\bibinfo {author} {\bibfnamefont {T.}~\bibnamefont
  {Giamarchi}},\ }\href@noop {} {\emph {\bibinfo {title} {Quantum Physics in
  One Dimension}}}\ (\bibinfo  {publisher} {Oxford},\ \bibinfo {address} {New
  Dehli},\ \bibinfo {year} {2004})\BibitemShut {NoStop}%
\bibitem [{\citenamefont {Deshpande}\ \emph {et~al.}(2010)\citenamefont
  {Deshpande}, \citenamefont {Bockrath}, \citenamefont {Glazman},\ and\
  \citenamefont {Yacoby}}]{Deshpande2010ela}%
  \BibitemOpen
  \bibfield  {author} {\bibinfo {author} {\bibfnamefont {V.~V.}\ \bibnamefont
  {Deshpande}}, \bibinfo {author} {\bibfnamefont {M.}~\bibnamefont {Bockrath}},
  \bibinfo {author} {\bibfnamefont {L.~I.}\ \bibnamefont {Glazman}}, \ and\
  \bibinfo {author} {\bibfnamefont {A.}~\bibnamefont {Yacoby}},\ }\Doi
  {10.1038/nature08918} {\bibfield  {journal} {\bibinfo  {journal} {Nature},\
  }\textbf {\bibinfo {volume} {464}},\ \bibinfo {pages} {209} (\bibinfo {year}
  {2010})}\BibitemShut {NoStop}%
\bibitem [{\citenamefont {Kane}\ and\ \citenamefont
  {Fisher}(1992)}]{Kane1992tia}%
  \BibitemOpen
  \bibfield  {author} {\bibinfo {author} {\bibfnamefont {C.~L.}\ \bibnamefont
  {Kane}}\ and\ \bibinfo {author} {\bibfnamefont {M.~P.~A.}\ \bibnamefont
  {Fisher}},\ }\Doi {10.1103/PhysRevB.46.15233} {\bibfield  {journal} {\bibinfo
   {journal} {Phys. Rev. B},\ }\textbf {\bibinfo {volume} {46}},\ \bibinfo
  {pages} {15233} (\bibinfo {year} {1992})}\BibitemShut {NoStop}%
\bibitem [{\citenamefont {Sassetti}\ and\ \citenamefont
  {Weiss}(1994)}]{Sassetti1994to1}%
  \BibitemOpen
  \bibfield  {author} {\bibinfo {author} {\bibfnamefont {M.}~\bibnamefont
  {Sassetti}}\ and\ \bibinfo {author} {\bibfnamefont {U.}~\bibnamefont
  {Weiss}},\ }\Doi {10.1209/0295-5075/27/4/010} {\bibfield  {journal} {\bibinfo
   {journal} {Europhys. Lett.},\ }\textbf {\bibinfo {volume} {27}},\ \bibinfo
  {pages} {311} (\bibinfo {year} {1994})}\BibitemShut {NoStop}%
\bibitem [{\citenamefont {Balents}(1999)}]{Balents1999oci}%
  \BibitemOpen
  \bibfield  {author} {\bibinfo {author} {\bibfnamefont {L.}~\bibnamefont
  {Balents}},\ }in\ \href@noop {} {\emph {\bibinfo {booktitle} {XVIII Moriond
  Les Arcs Conference Proceedings}}},\ \bibinfo {editor} {edited by\ \bibinfo
  {editor} {\bibfnamefont {D.~C.}\ \bibnamefont {Glattli}}\ and\ \bibinfo
  {editor} {\bibfnamefont {M.}~\bibnamefont {Sanquer}}}\ (\bibinfo  {publisher}
  {Edition Frontiers},\ \bibinfo {address} {Paris},\ \bibinfo {year} {1999})\
  \Eprint {http://arxiv.org/abs/cond-mat/9906032} {; arXiv:cond-mat/9906032}
  \BibitemShut {NoStop}%
\bibitem [{\citenamefont {Mishchenko}\ \emph {et~al.}(2001)\citenamefont
  {Mishchenko}, \citenamefont {Andreev},\ and\ \citenamefont
  {Glazman}}]{Mishchenko2001zba}%
  \BibitemOpen
  \bibfield  {author} {\bibinfo {author} {\bibfnamefont {E.~G.}\ \bibnamefont
  {Mishchenko}}, \bibinfo {author} {\bibfnamefont {A.~V.}\ \bibnamefont
  {Andreev}}, \ and\ \bibinfo {author} {\bibfnamefont {L.~I.}\ \bibnamefont
  {Glazman}},\ }\Doi {10.1103/PhysRevLett.87.246801} {\bibfield  {journal}
  {\bibinfo  {journal} {Phys. Rev. Lett.},\ }\textbf {\bibinfo {volume} {87}},\
  \bibinfo {pages} {246801} (\bibinfo {year} {2001})}\BibitemShut {NoStop}%
\bibitem [{\citenamefont {Tartakovskii}\ \emph {et~al.}(1987)\citenamefont
  {Tartakovskii}, \citenamefont {Fistul'}, \citenamefont {Raikh},\ and\
  \citenamefont {Ruzin}}]{Tartakovskii1987hco}%
  \BibitemOpen
  \bibfield  {author} {\bibinfo {author} {\bibfnamefont {A.~V.}\ \bibnamefont
  {Tartakovskii}}, \bibinfo {author} {\bibfnamefont {M.~V.}\ \bibnamefont
  {Fistul'}}, \bibinfo {author} {\bibfnamefont {M.~E.}\ \bibnamefont {Raikh}},
  \ and\ \bibinfo {author} {\bibfnamefont {I.~M.}\ \bibnamefont {Ruzin}},\
  }\href@noop {} {\bibfield  {journal} {\bibinfo  {journal} {Sov. Phys.
  Semicond.},\ }\textbf {\bibinfo {volume} {21}},\ \bibinfo {pages} {603}
  (\bibinfo {year} {1987})}\BibitemShut {NoStop}%
\bibitem [{\citenamefont {Levin}\ \emph {et~al.}(1988)\citenamefont {Levin},
  \citenamefont {Ruzin},\ and\ \citenamefont {Shklovskii}}]{Levin1988thc}%
  \BibitemOpen
  \bibfield  {author} {\bibinfo {author} {\bibfnamefont {E.~I.}\ \bibnamefont
  {Levin}}, \bibinfo {author} {\bibfnamefont {I.~M.}\ \bibnamefont {Ruzin}}, \
  and\ \bibinfo {author} {\bibfnamefont {B.~I.}\ \bibnamefont {Shklovskii}},\
  }\href@noop {} {\bibfield  {journal} {\bibinfo  {journal} {Sov. Phys.
  Semicond.},\ }\textbf {\bibinfo {volume} {22}},\ \bibinfo {pages} {401}
  (\bibinfo {year} {1988})}\BibitemShut {NoStop}%
\bibitem [{\citenamefont {Glazman}\ and\ \citenamefont
  {Matveev}(1988)}]{Glazman1988ita}%
  \BibitemOpen
  \bibfield  {author} {\bibinfo {author} {\bibfnamefont {L.~I.}\ \bibnamefont
  {Glazman}}\ and\ \bibinfo {author} {\bibfnamefont {K.~A.}\ \bibnamefont
  {Matveev}},\ }\href@noop {} {\bibfield  {journal} {\bibinfo  {journal} {Sov.
  Phys. JETP},\ }\textbf {\bibinfo {volume} {67}},\ \bibinfo {pages} {332}
  (\bibinfo {year} {1988})}\BibitemShut {NoStop}%
\bibitem [{\citenamefont {Bahlouli}\ \emph {et~al.}(1994)\citenamefont
  {Bahlouli}, \citenamefont {Matveev}, \citenamefont {Ephron},\ and\
  \citenamefont {Beasley}}]{Bahlouli1994cci}%
  \BibitemOpen
  \bibfield  {author} {\bibinfo {author} {\bibfnamefont {H.}~\bibnamefont
  {Bahlouli}}, \bibinfo {author} {\bibfnamefont {K.~A.}\ \bibnamefont
  {Matveev}}, \bibinfo {author} {\bibfnamefont {D.}~\bibnamefont {Ephron}}, \
  and\ \bibinfo {author} {\bibfnamefont {M.~R.}\ \bibnamefont {Beasley}},\
  }\Doi {10.1103/PhysRevB.49.14496} {\bibfield  {journal} {\bibinfo  {journal}
  {Phys. Rev. B},\ }\textbf {\bibinfo {volume} {49}},\ \bibinfo {pages} {14496}
  (\bibinfo {year} {1994})}\BibitemShut {NoStop}%
\bibitem [{\citenamefont {Rodin}\ and\ \citenamefont
  {Fogler}(2009)}]{Rodin2009nso}%
  \BibitemOpen
  \bibfield  {author} {\bibinfo {author} {\bibfnamefont {A.~S.}\ \bibnamefont
  {Rodin}}\ and\ \bibinfo {author} {\bibfnamefont {M.~M.}\ \bibnamefont
  {Fogler}},\ }\Doi {10.1103/PhysRevB.80.155435} {\bibfield  {journal}
  {\bibinfo  {journal} {Phys. Rev. B},\ }\textbf {\bibinfo {volume} {80}},\
  \bibinfo {pages} {155435} (\bibinfo {year} {2009})}\BibitemShut {NoStop}%
\bibitem [{\citenamefont {Pollak}\ and\ \citenamefont
  {Riess}(1976)}]{Pollak1976apt}%
  \BibitemOpen
  \bibfield  {author} {\bibinfo {author} {\bibfnamefont {M.}~\bibnamefont
  {Pollak}}\ and\ \bibinfo {author} {\bibfnamefont {I.}~\bibnamefont {Riess}},\
  }\Doi {10.1088/0022-3719/9/12/017} {\bibfield  {journal} {\bibinfo  {journal}
  {J. Phys. C},\ }\textbf {\bibinfo {volume} {9}},\ \bibinfo {pages} {2339}
  (\bibinfo {year} {1976})}\BibitemShut {NoStop}%
\bibitem [{\citenamefont {Miller}\ and\ \citenamefont
  {Abrahams}(1960)}]{Miller1960ica}%
  \BibitemOpen
  \bibfield  {author} {\bibinfo {author} {\bibfnamefont {A.}~\bibnamefont
  {Miller}}\ and\ \bibinfo {author} {\bibfnamefont {E.}~\bibnamefont
  {Abrahams}},\ }\Doi {10.1103/PhysRev.120.745} {\bibfield  {journal} {\bibinfo
   {journal} {Phys. Rev.},\ }\textbf {\bibinfo {volume} {120}},\ \bibinfo
  {pages} {745} (\bibinfo {year} {1960})}\BibitemShut {NoStop}%
\bibitem [{\citenamefont {Ballard}\ and\ \citenamefont
  {Raikh}(2006)}]{Ballard2006roc}%
  \BibitemOpen
  \bibfield  {author} {\bibinfo {author} {\bibfnamefont {A.~D.}\ \bibnamefont
  {Ballard}}\ and\ \bibinfo {author} {\bibfnamefont {M.~E.}\ \bibnamefont
  {Raikh}},\ }\Doi {10.1103/PhysRevB.74.035117} {\bibfield  {journal} {\bibinfo
   {journal} {Phys. Rev. B},\ }\textbf {\bibinfo {volume} {74}},\ \bibinfo
  {pages} {035117} (\bibinfo {year} {2006})}\BibitemShut {NoStop}%
\bibitem [{\citenamefont {Park}(2000)}]{Park2000thc}%
  \BibitemOpen
  \bibfield  {author} {\bibinfo {author} {\bibfnamefont {Y.}~\bibnamefont
  {Park}},\ }\Doi {10.1016/S0038-1098(00)00196-4} {\bibfield  {journal}
  {\bibinfo  {journal} {Solid State Commun.},\ }\textbf {\bibinfo {volume}
  {115}},\ \bibinfo {pages} {281} (\bibinfo {year} {2000})}\BibitemShut
  {NoStop}%
\bibitem [{\citenamefont {Xu}\ \emph {et~al.}(1995)\citenamefont {Xu},
  \citenamefont {Ephron},\ and\ \citenamefont {Beasley}}]{Xu1995dih}%
  \BibitemOpen
  \bibfield  {author} {\bibinfo {author} {\bibfnamefont {Y.}~\bibnamefont
  {Xu}}, \bibinfo {author} {\bibfnamefont {D.}~\bibnamefont {Ephron}}, \ and\
  \bibinfo {author} {\bibfnamefont {M.~R.}\ \bibnamefont {Beasley}},\ }\Doi
  {10.1103/PhysRevB.52.2843} {\bibfield  {journal} {\bibinfo  {journal} {Phys.
  Rev. B},\ }\textbf {\bibinfo {volume} {52}},\ \bibinfo {pages} {2843}
  (\bibinfo {year} {1995})}\BibitemShut {NoStop}%
\bibitem [{\citenamefont {Kurkij\"{a}rvi}(1973)}]{Kurkijarvi1973hci}%
  \BibitemOpen
  \bibfield  {author} {\bibinfo {author} {\bibfnamefont {J.}~\bibnamefont
  {Kurkij\"{a}rvi}},\ }\Doi {10.1103/PhysRevB.8.922} {\bibfield  {journal}
  {\bibinfo  {journal} {Phys. Rev. B},\ }\textbf {\bibinfo {volume} {8}},\
  \bibinfo {pages} {922} (\bibinfo {year} {1973})}\BibitemShut {NoStop}%
\bibitem [{\citenamefont {Lee}(1984)}]{Lee1984vrh}%
  \BibitemOpen
  \bibfield  {author} {\bibinfo {author} {\bibfnamefont {P.~A.}\ \bibnamefont
  {Lee}},\ }\Doi {10.1103/PhysRevLett.53.2042} {\bibfield  {journal} {\bibinfo
  {journal} {Phys. Rev. Lett.},\ }\textbf {\bibinfo {volume} {53}},\ \bibinfo
  {pages} {2042} (\bibinfo {year} {1984})}\BibitemShut {NoStop}%
\bibitem [{\citenamefont {Raikh}\ and\ \citenamefont
  {Ruzin}(1989)}]{Raikh1989fot}%
  \BibitemOpen
  \bibfield  {author} {\bibinfo {author} {\bibfnamefont {M.~E.}\ \bibnamefont
  {Raikh}}\ and\ \bibinfo {author} {\bibfnamefont {I.~M.}\ \bibnamefont
  {Ruzin}},\ }\href@noop {} {\bibfield  {journal} {\bibinfo  {journal} {Sov.
  Phys. JETP},\ }\textbf {\bibinfo {volume} {68}},\ \bibinfo {pages} {1113}
  (\bibinfo {year} {1989})}\BibitemShut {NoStop}%
\bibitem [{\citenamefont {Fogler}\ and\ \citenamefont
  {Kelley}(2005)}]{Fogler2005nov}%
  \BibitemOpen
  \bibfield  {author} {\bibinfo {author} {\bibfnamefont {M.~M.}\ \bibnamefont
  {Fogler}}\ and\ \bibinfo {author} {\bibfnamefont {R.~S.}\ \bibnamefont
  {Kelley}},\ }\Doi {10.1103/PhysRevLett.95.166604} {\bibfield  {journal}
  {\bibinfo  {journal} {Phys. Rev. Lett.},\ }\textbf {\bibinfo {volume} {95}},\
  \bibinfo {pages} {166604} (\bibinfo {year} {2005})}\BibitemShut {NoStop}%
\bibitem [{\citenamefont {Beenakker}(1997)}]{Beenakker1997rmt}%
  \BibitemOpen
  \bibfield  {author} {\bibinfo {author} {\bibfnamefont {C.~W.~J.}\
  \bibnamefont {Beenakker}},\ }\Doi {10.1103/RevModPhys.69.731} {\bibfield
  {journal} {\bibinfo  {journal} {Rev. Mod. Phys.},\ }\textbf {\bibinfo
  {volume} {69}},\ \bibinfo {pages} {731} (\bibinfo {year} {1997})}\BibitemShut
  {NoStop}%
\bibitem [{\citenamefont {Fogler}\ \emph {et~al.}(2004)\citenamefont {Fogler},
  \citenamefont {Teber},\ and\ \citenamefont {Shklovskii}}]{Fogler2004vrh}%
  \BibitemOpen
  \bibfield  {author} {\bibinfo {author} {\bibfnamefont {M.~M.}\ \bibnamefont
  {Fogler}}, \bibinfo {author} {\bibfnamefont {S.}~\bibnamefont {Teber}}, \
  and\ \bibinfo {author} {\bibfnamefont {B.~I.}\ \bibnamefont {Shklovskii}},\
  }\Doi {10.1103/PhysRevB.69.035413} {\bibfield  {journal} {\bibinfo  {journal}
  {Phys. Rev. B},\ }\textbf {\bibinfo {volume} {69}},\ \bibinfo {pages}
  {035413} (\bibinfo {year} {2004})}\BibitemShut {NoStop}%
\bibitem [{\citenamefont {Fogler}\ \emph {et~al.}(2006)\citenamefont {Fogler},
  \citenamefont {Malinin},\ and\ \citenamefont
  {N\"{a}ttermann}}]{Fogler2006cba}%
  \BibitemOpen
  \bibfield  {author} {\bibinfo {author} {\bibfnamefont {M.~M.}\ \bibnamefont
  {Fogler}}, \bibinfo {author} {\bibfnamefont {S.~V.}\ \bibnamefont {Malinin}},
  \ and\ \bibinfo {author} {\bibfnamefont {T.}~\bibnamefont {N\"{a}ttermann}},\
  }\Doi {10.1103/PhysRevLett.97.096601} {\bibfield  {journal} {\bibinfo
  {journal} {Phys. Rev. Lett.},\ }\textbf {\bibinfo {volume} {97}},\ \bibinfo
  {pages} {096601} (\bibinfo {year} {2006})}\BibitemShut {NoStop}%
\bibitem [{\citenamefont {Deroulers}()}]{Deroulers2007dot}%
  \BibitemOpen
  \bibfield  {author} {\bibinfo {author} {\bibfnamefont {C.}~\bibnamefont
  {Deroulers}},\ }\href@noop {} {\enquote {\bibinfo {title} {Distribution of
  the resistance of nanowires with strong impurities},}\ }\Eprint
  {http://arxiv.org/abs/0705.3090} {arXiv:0705.3090} \BibitemShut {NoStop}%
\bibitem [{\citenamefont {Tran}\ \emph {et~al.}(2005)\citenamefont {Tran},
  \citenamefont {Beloborodov}, \citenamefont {Lin}, \citenamefont {Bigioni},
  \citenamefont {Vinokur},\ and\ \citenamefont {Jaeger}}]{Tran2005mci}%
  \BibitemOpen
  \bibfield  {author} {\bibinfo {author} {\bibfnamefont {T.~B.}\ \bibnamefont
  {Tran}}, \bibinfo {author} {\bibfnamefont {I.~S.}\ \bibnamefont
  {Beloborodov}}, \bibinfo {author} {\bibfnamefont {X.~M.}\ \bibnamefont
  {Lin}}, \bibinfo {author} {\bibfnamefont {T.~P.}\ \bibnamefont {Bigioni}},
  \bibinfo {author} {\bibfnamefont {V.~M.}\ \bibnamefont {Vinokur}}, \ and\
  \bibinfo {author} {\bibfnamefont {H.~M.}\ \bibnamefont {Jaeger}},\ }\Doi
  {10.1103/PhysRevLett.95.076806} {\bibfield  {journal} {\bibinfo  {journal}
  {Phys. Rev. Lett.},\ }\textbf {\bibinfo {volume} {95}},\ \bibinfo {pages}
  {076806} (\bibinfo {year} {2005})}\BibitemShut {NoStop}%
\end{thebibliography}

%

\end{document}